\documentclass[12pt]{iopart}

\newcommand{\etalcite}[2]
{#1 \emph{et al.}~\cite{#2}}

\newcommand{\subsize}
{\scriptsize}

\newcommand{\subsubsize}
{\tiny}

\newcommand{\sub}[1]
{_\textrm{\subsize #1}}

\newcommand{\subsub}[1]
{_\textrm{\subsubsize #1}}

\newcommand{\gammafkt}[1]
{\Gamma\big(\frac{#1}{2}\big)}

\newcommand{\gammafktB}[2]
{\frac{\Gamma\big(\frac{#1}{2}\big)}{\Gamma\big(\frac{#2}{2}\big)}}

\newcommand{\BE}[1]
{Bose-Einstein}

\usepackage{iopams}  
\usepackage{varioref}

\begin{document}

\title{Structure of boson systems beyond the mean-field}

\author{O.~S{\o}rensen \footnote[3]{To whom correspondence should be
    addressed (oles@phys.au.dk)}, D.~V.~Fedorov, and A.~S.~Jensen}

\address{Department of Physics and Astronomy, University of Aarhus,
  DK-8000 Aarhus C, Denmark}

\begin{abstract} 
  We investigate systems of identical bosons with the focus on
  two-body correlations.  We use the hyperspherical adiabatic method
  and a decomposition of the wave function in two-body amplitudes.  An
  analytic parametrization is used for the adiabatic effective radial
  potential.  We discuss the structure of a condensate for arbitrary
  scattering length. Stability and time scales for various decay
  processes are estimated.  The previously predicted Efimov-like
  states are found to be very narrow. We discuss the validity
  conditions and formal connections between the zero- and finite-range
  mean-field approximations, Faddeev-Yakubovski{\u\i} formulation,
  Jastrow ansatz, and the present method.  We compare numerical
  results from present work with mean-field calculations and discuss
  qualitatively the connection with measurements.
\end{abstract}


\submitto{\JPB}

\date{today}

\section{Introduction}

The interest in dilute Bose gases has been growing since the
experimental realisation of the phenomenon of Bose-Einstein
condensation (BEC) \cite{and95,bra95,dav95}.  Excellent reviews of the
world of BEC are given in recently published monographs
\cite{pet01,pit03}.  The theoretical interest in BEC goes more than
fifty years back and is widely based on the mean-field formulation.
The usual measure of the validity of the mean-field is that $n|a_s|^3
\ll 1$, where $n$ is the density and $a_s$ is the two-body $s$-wave
scattering length \cite{pet01}.  Then the particles are not too close
to each other and correlations are expected to be negligibly small.
The importance of correlations must increase with the density of the
system and the mean-field method sooner or later becomes inadequate.

A Feshbach resonance is routinely used to create Bose-Einstein
condensed systems, where the effective interaction corresponds to a
large scattering length $a_s$ \cite{tim99}.  Then stronger correlated
structures arise and the condensate becomes unstable as seen
experimentally \cite{don01}.  A theoretical description based on the
time-dependent Gross-Pitaevskii equation (zero-range two-body
interaction) was given in \cite{adh02b,adh02d}.  By definition then
only average properties are incorporated although the dominating decay
mechanism is very sensitive to correlations.

Correlations are clearly fully included in exact solutions of the full
problem as attempted in few-body physics. For fermionic systems the
current limit is about 10 particles, see e.g., \cite{pie02}.  The
simplifications for identical bosons allow computations of a larger
number of particles especially when using variational methods like the
quantum Monte Carlo method, see e.g., \cite{bre01}. However, detailed
investigations for relatively few boson or fermion systems already
require a substantial effort \cite{blu02c,ast03}.  A larger number of
particles can be handled if only specific properties are wanted and
not the full correlated wave function. In particular quantum Monte
Carlo calculations have reproduced density profiles in agreement with
the Gross-Pitaevskii results \cite{kra96,cer00}.

The need to account for correlations seems unavoidable in experiments
where cluster structure is important, e.g., formation of molecular
dimer states.  The simplest example is probably the three-body
recombination process, where two of the atoms in the many-body system
react and form a two-body bound state.  In \cite{tim99b} is suggested
to apply a Feshbach resonance to create a hybrid atomic-molecular
Bose-Einstein condensate.  The atom-molecule coupling is included on
top of the usual mean-field equations.  It is then conjectured, that
the ground state of a Feshbach resonant Bose-Einstein condensate in
reality is a mixed condensate of atoms and di-atomic molecules.
Adhikari \cite{adh01b} studied the coupled system of atoms and
molecules from the Gross-Pitaevskii equation and predicted oscillation
phenomena.  The experiment reported in \cite{don02} used the tuning of
a Feshbach resonance with a collapse-burst process as a result.  It
was subsequently shown \cite{kok02b,mac02,koh03b} that the coherent
burst-remnant oscillations could be accounted for by the presence of a
molecular Bose-Einstein condensate.  Furthermore, the creation of a
di-atomic molecular condensate of fermionic atoms was recently
observed \cite{reg03,str03} giving additional evidence for the
creation of a mixed condensate.

Descriptions of correlations within $N$-body systems suggest the use
of few-body techniques which are tools to understand the few-body
structures essentially decoupled from all other degrees of freedom.
This suggests to extend the use of suitable three-body formulations.
A particularly promising set of calculations were reported in
\cite{blu02c} for an isolated three-body system with total angular
momentum zero.  They varied the scattering length and described
systems with any number of bound two-body states.  Moreover, they
studied a range of excited states and concluded that such higher-lying
condensate-like states do not collapse under the usual conditions when
$N|a_s|/b\sub t>0.5$ \cite{rob01}, where $b\sub t$ is the trap length
of an external harmonic field.

Generalization of this work to $N$-body systems started in
\cite{boh98}, where average properties of boson systems were
investigated with hyperspherical coordinates.  In \cite{sor02,sor02b}
this adiabatic hyperspherical method was extended to explicitly
include two-body correlations in $N$-body boson systems.  This
structure of the wave function is beyond the mean-field.  The
application for the particle number $N=20$ was extended up to $N=10^5$
\cite{sor03a}.  Scaling properties were deduced as function of
scattering length and particle number and analytic expressions were
derived for the adiabatic potentials.

The results from \cite{sor02,sor02b,sor03a} indicate that the
mean-field properties for dilute systems are reproduced in addition to
the two-body correlations.  Better understandings of validity
conditions and connections between mean-field models and the adiabatic
hyperspherical expansion method are desirable.  Of particular interest
are correlations for large scattering lengths, which cannot be studied
by mean-field methods.

Before proceeding it may be useful to express our definition of
correlations in $N$-body systems, i.e., as structures indescribable by
mean-field wave functions.  For example, if two particles form a bound
state it is possible to formulate a mean-field theory for such dimer
states. It is also possible to construct a theoretical formulation
where mean-field wave functions are used for each species in a coupled
system of single particles and dimers. Even when dimers can separate
and two particles combine to dimers this could still appropriately be
called a mean-field treatment.  Examples are the
Hartree-Fock-Bogoliubov (HFB) formulation in \cite{kok02b} for atoms
and molecular (bound or unbound) dimers and the HFB approximation for
nucleons and unbound pairs of nucleons frequently applied to the
nuclear many-body problem \cite{dea03}.  When all divergences are
removed by renormalization, the restriction of the Hilbert space to
mean-field wave functions allow computations for any interaction
parameter, e.g., large two-body scattering length.  Obviously, this
does not imply that the true many-body correlations can be described,
only the structures allowed by the wave function.  The key point is
the allowed Hilbert space in a specific formulation.  In \cite{kok02b}
both atoms and molecular dimers are simultaneously allowed.  In
\cite{sor02,sor02b,sor03a} not only molecular dimers are allowed, but
all kinds of diatomic correlated structures with non-zero higher-order
correlation functions are included.

The purpose of this article is to discuss both qualitative and
quantitative gross properties of $N$-body boson systems where two-body
correlations explicitly are included.  In section~\ref{sec:theory} we
briefly summarize the hyperspherical theory for studying correlations.
The connections to mean-field methods and other descriptions of
correlations in many-body systems are not previously formulated and we
include a general discussion in section~\ref{sec:wave-function}.  In
section~\ref{sec:general-properties} we present the hyperspherical
potentials and discuss the analytic parametrization of the effective
radial potential extracted in \cite{sor03a}. We derive scaling
properties and discuss qualitatively a possible scenario for decay and
collapse of the condensate after sudden changes of the effective
two-body interactions.  In section~\ref{sec:conn-mean-field} we
discuss details of relations to the mean-field, improvements over the
mean-field, and ranges of validity of the mean-field and the present
hyperspherical method.  This section contains essentially only new
results.  Finally, we summarize and conclude in
section~\ref{sec:summary-conclusion}.

\section{Theory}

\label{sec:theory}

We study $N$ identical bosons of mass $m$ trapped by an external
harmonic field of angular frequency $\omega$.  We assume interaction
via a short-range two-body potential $V$, which may depend on the
relative spin state. However, we shall see that the interaction
basically only enters through the relevant scattering length and the
formulation is therefore the same as for spinless bosons.  The
Hamiltonian is then given by
\begin{eqnarray}
  \hat H
  &=&
  \sum_{i=1}^N\Big(\frac{\hat p_i^2}{2m}+\frac12m\omega^2r_i^2\Big)+
  \sum_{i<j}^NV(r_{ij}) 
  \;,
  \label{eq:Hamiltonian}
\end{eqnarray}
where $\vec r_i$ is the position of particle $i$, $\vec p_i$ the
conjugate momentum, and $r_{ij} = |\vec r_j-\vec r_i|$ is the
interparticle distance.  The interaction part is independent of the
center of mass.  Both the kinetic energy and the external harmonic
field can be separated into parts depending on the center of mass and
parts only depending on relative coordinates. For this we use the
relation
\begin{eqnarray}
  \sum_{i=1}^Nr_i^2
  =
  \frac1N\sum_{i<j}^Nr_{ij}^2
  +NR^2
  \;,
\end{eqnarray}
where $\vec R = \sum_i\vec r_i/N$ is the center of mass coordinate.
This immediately leads to the convenient definition of the hyperradius
$\rho$
\begin{eqnarray} \label{e3}
  \rho^2
  &\equiv&
  \frac1N\sum_{i<j}^Nr_{ij}^2=\sum_{i=1}^Nr_i^2-NR^2
  \;.
\end{eqnarray}
The relative degrees of freedom are first related by $N-1$ Jacobi
vectors $\vec\eta_k$ \cite{bar99b}.  We next choose a new set of
coordinates, the hyperspherical coordinates, to describe the $3N-3$
relative degrees of freedom.  The hyperradius $\rho$ sets the overall
length scale for the system, the angles $\alpha_k$ determine the $N-2$
relations $\eta_k = \rho \cos\alpha_{N-1} \cos\alpha_{N-2} \ldots
\cos\alpha_{k+1} \sin\alpha_k$ between the lengths of the Jacobi
vectors, and $2(N-1)$ angles determine the orientations of the Jacobi
vectors.  All $3N-4$ hyperangles are collectively denoted by $\Omega$
\cite{bar99a,sor01}.

\subsection{Adiabatic hyperspherical method}

The Hamiltonian then separates into a center of mass part, a radial
part, and an angular part depending respectively on $\vec R$, $\rho$,
and $\Omega$ \cite{sor01}
\begin{eqnarray}
  \hat H 
  &=&
  \hat H\sub{c.m.}
  +
  \hat H_\rho+\frac{\hbar^2\hat h_\Omega}{2m\rho^2}
  \;,
  \\
  \hat H\sub{c.m.}
  &=&
  \frac{\hat p_R^2}{2Nm}+\frac12Nm\omega^2R^2
  \;,
  \\
  \hat H_\rho
  &=&
  -\frac{\hbar^2}{2m}\frac{1}{\rho^{3N-4}}\frac{\partial}{\partial\rho}
  \rho^{3N-4}\frac{\partial}{\partial\rho}
  +\frac12m\omega^2\rho^2
  \;,
  \\
  \hat h_\Omega
  &=&
  \hat\Lambda_{N-1}^2+\frac{2m\rho^2}{\hbar^2}\sum_{i<j}^NV(r_{ij})
  \;.
\end{eqnarray}
The angular kinetic energy operator $\hat\Lambda_{N-1}^2$ is given
recursively by
\begin{eqnarray}
  \hat\Lambda_k^2
  =\hat\Pi_k^2+
  \frac{\hat\Lambda_{k-1}^2}{\cos^2\alpha_k}
  +\frac{\hat l_k^2}{\sin^2\alpha_k}
  \;,\quad
  \\
  \hat\Pi_k^2
  =-\frac{\partial^2}{\partial\alpha_k^2}+
  \frac{3k-6-(3k-2)\cos 2\alpha_k}{\sin 2 \alpha_k}
  \frac{\partial}{\partial\alpha_k}
  \;,
\end{eqnarray}
where $\hat l_k$ is the angular momentum operator associated with
$\vec\eta_k$.

The relative wave function $\Psi(\rho,\Omega)$ obeys the Schr\"odinger
equation
\begin{eqnarray}
  (\hat H-\hat H\sub{c.m.})\Psi(\rho,\Omega)
  =
  E\Psi(\rho,\Omega)
  \;.
\end{eqnarray}
The adiabatic expansion of the wave function is
\begin{eqnarray}
  \Psi(\rho,\Omega)
  &=&
  \sum_{\nu=0}^\infty
  F_{\nu}(\rho)\Phi_\nu(\rho,\Omega)
  \;,\quad
  F_{\nu}(\rho)
  =
  \rho^{-(3N-4)/2}
  f_{\nu}(\rho)
  \;,
  \label{eq:hyperspherical_wave_function}
\end{eqnarray}
where $\Phi_\nu$ is an eigenfunction of the angular part of the
Hamiltonian with an eigenvalue $\hbar^2\lambda_\nu(\rho)/(2m\rho^2)$
\begin{eqnarray}
  \hat h_\Omega\Phi_\nu(\rho,\Omega)
  &=&
  \lambda_\nu(\rho)\Phi_\nu(\rho,\Omega)
  \;.
  \label{eq:angular_equation}
\end{eqnarray}
Neglecting couplings between the different $\nu$-channels yields the
radial eigenvalue equation for the eigenfunction $f_\nu$ and the
energy $E_\nu$
\begin{eqnarray}
  &&
  \Big(-\frac{\hbar^2}{2m}\frac{d^2}{d\rho^2} + U_\nu(\rho) - E_{\nu}\Big)
  f_{\nu}(\rho)
  =
  0
  \;,
  \label{eq:radial.equation}
  \\
  &&
  \frac{2mU_\nu(\rho)}{\hbar^2}
  =
  \frac{\lambda_\nu}{\rho^2}+
  \frac{(3N-4)(3N-6)}{4\rho^2}+
  \frac{\rho^2}{b\sub t^4}
  \;,
  \label{eq:radial.potential}
\end{eqnarray}
where $b\sub t \equiv \sqrt{\hbar/(m\omega)}$ is the trap length and
the adiabatic potential $U_\nu$ is a function of the hyperradius. It
consists of three terms, i.e., the external field, the generalized
centrifugal barrier, and the angular average of the interactions and
kinetic energies.  The neglected non-diagonal terms are for large
hyperradii less than 1\% of the diagonal terms for attractive Gaussian
potentials.  Thus, the center of mass motion is separated out and the
hyperspherical adiabatic method is promising simply due to small
coupling terms.  The remaining problem is the determination of the
angular potential $\lambda$ from the angular eigenvalue equation.

\subsection{The wave function}

\label{sec:wave-function}

We have so far not assumed specific structures or restricted the
allowed Hilbert space for the many-body wave function.  At some point
we need to make a suitable ansatz for the angular wave function
$\Phi_{\nu}(\rho,\Omega)$.  However, first we shall relate to the
historically successful approaches to describe a many-body wave
function.

\subsubsection{The Hartree mean-field description.}

\label{sec:mean-field-descr}

The ground-state Hartree product of single-particle amplitudes
\cite{bra83}
\begin{eqnarray}
  \Psi\sub{H}(\vec r_1,\vec r_2,\ldots,\vec r_N)
  =
  \prod_{i=1}^N
  \psi\sub{s.p.}(\vec r_i)
  \;,
\end{eqnarray}
is for the non-interacting gas in the external field given by the
amplitude
\begin{eqnarray}
  \psi\sub{s.p.}(\vec r_i)
  =
  C
  e^{-r_i^2/(2b\sub t^2)}
  \;,\quad
  C^{-1}=\pi^{3/4}b\sub t^{3/2}
  \;.
\end{eqnarray}
With the relation $\sum_{i=1}^N r_i^2=\rho^2+NR^2$ this can be
rewritten as
\begin{eqnarray}
  \Psi\sub{H}(\vec r_1,\vec r_2,\ldots,\vec r_N)
  =
  C^N
  \exp\bigg[-\sum_{i=1}^N r_i^2/(2b\sub t^2)\bigg]
  \nonumber\\
  =
  C^N
  e^{-\rho^2/(2b\sub t^2)}e^{-NR^2/(2b\sub t^2)}
  =
  \Upsilon_0(\vec R)F_0(\rho)\Phi_0
  \;.
  \label{eq:Gauss_product}
\end{eqnarray}
The separation of the center of mass motion assures that the
ground-state center of mass function always is $\Upsilon_0(\vec R) =
CN^{3/4} \exp[-NR^2/(2b\sub t^2)]$.  Then
equation~(\ref{eq:Gauss_product}) is a product of the lowest solution
for the motion of the center of mass in a trap, and the lowest
hyperspherical wave function $F_0 \Phi_0$ in
equation~(\ref{eq:hyperspherical_wave_function}), where $F_0(\rho)
\propto \exp[-\rho^2/(2b\sub t^2)]$ and the angular part
$\Phi_0(\rho,\Omega)$ is a constant.  This relation in
equation~(\ref{eq:Gauss_product}) between ordinary cartesian and
hyperspherical coordinates is valid for any length parameter $b\sub
t$.  Therefore a mean-field product of identical single-particle
Gaussian wave functions is equivalent to a hyperradial Gaussian and a
constant angular wave function, i.e., with no dependence on
hyperangles $\Omega$.

In reality the interactions produce correlations and the
hyperspherical wave function deviates from a hyperradial Gaussian
multiplied by a constant hyperangular part.  Therefore the mean-field
Hartree product wave function is not exact. However, a measure can be
obtained by calculating the single-particle density $n$, given by
\begin{eqnarray} \label{e16}
  n(\vec r_1)
  =
  \int d^3\vec r_2 d^3\vec r_3\cdots d^3\vec r_N
  |\Psi(\rho,\Omega)\Upsilon_0(\vec R)|^2
  \;,
\end{eqnarray}
which can be compared with the mean-field analogue
$|\psi\sub{s.p.}(\vec r_1)|^2$.  The $3(N-1)$-dimensional integral in
equation~(\ref{e16}) is very complicated with the full numerical
hyperspherical solution.  To get an idea of the possible structures we
assume a constant angular part $\Phi(\rho,\Omega)$.  We expand the
hyperradial density distribution on Gaussian amplitudes with different
length parameters $a_j$:
\begin{eqnarray}
  |F(\rho)|^2
  =
  \sum_j
  c_j\;
  \frac{2}{\gammafkt{3N-3}a_j^{3N-3}}
  e^{-\rho^2/a_j^2}
  \;,
\end{eqnarray}
where $\sum_j c_j=1$ assures that $F(\rho)$ is normalized as
$\int_0^\infty d\rho \rho^{3N-4} |F(\rho)|^2=1$.  This yields
\begin{eqnarray}
  n(\vec r_1)
  =
  \sum_j 
  c_j\;
  \frac{1}{\pi^{3/2}B_j^3}
  e^{-r_1^2/B_j^2}
  \;,\quad
  B_j^2=\frac{(N-1)a_j^2+b\sub t^2}{N}
  \;,
\end{eqnarray}
which is equivalent to $\langle r_1^2\rangle = \int d^3\vec r_1\;
n(\vec r_1)r_1^2$, since 
\begin{eqnarray}
  \langle r_1^2\rangle
  =
  \frac{1}{N}\langle\rho^2\rangle+\langle R^2\rangle
  =
  \frac{3}{2}\Big(1-\frac{1}{N}\Big)\sum_j c_j a_j^2
  +\frac{3}{2}\frac{1}{N}b\sub{t}^2
\end{eqnarray}
and
\begin{eqnarray}
  &&
  \int d^3\vec r_1\; n(\vec r_1)r_1^2
  =
  \frac{3}{2}\sum_j c_jB_j^2
  \nonumber\\
  &&
  =
  \frac{3}{2}\Big(1-\frac{1}{N}\Big)\sum_j c_j a_j^2
  +\frac{3}{2}\frac{1}{N}b\sub{t}^2\sum_jc_j
  =
  \langle r_1^2\rangle
  \;.
\end{eqnarray}
The mean square distance between the particles can then be obtained
from the Gross-Pitaevskii, or mean-field, approximation for $\langle
r_1^2\rangle$ by the relation
\begin{eqnarray} \label{e21}
  \langle r_{12}^2\rangle
  =
  \frac{2N}{N-1}
  \Big(\langle r_1^2\rangle-\langle R^2\rangle\Big)
  =
  \frac{2N}{N-1}
  \Big(\langle r_1^2\rangle-\frac{1}{N}\frac{3}{2}b\sub t^2\Big)
  \;.
\end{eqnarray}
For a non-interacting gas in a harmonic external field the energy
$E_0$ is related to $E_0 \propto \langle r_1^2\rangle$ by the virial
theorem.

These relations are derived and valid only for Gaussian wave
functions.  However, the true mean-field solution is not strictly a
Gaussian, although such an approximation is rather efficient as
pointed out by \etalcite{Pethick}{pet01}.  The above results can be
used to relate an approximate Gaussian mean-field density distribution
to a similar hyperradial distribution implicitly, assuming constant
angular wave function corresponding to uncorrelated structure, see
\etalcite{Bohn}{boh98}.

\subsubsection{Faddeev-Yakubovski{\u\i} description.}

We seek the effect of correlations and have to operate beyond the
mean-field. Let us first consider the Faddeev-Yakubovski{\u\i}
techniques where the proper asymptotic behaviour of the wave functions
directly is taken into account \cite{fad61,yak67}. This formulation is
well suited when the large distance assymptotics is crucial as
expected for low-density systems.

Faddeev \cite{fad61} initially studied three-particle systems ($N=3$)
where one of the two-body subsystems is bound, and the other
subsystems are unbound.  He wrote the wave function as
$\Phi=\phi_{12}+\phi_{13} +\phi_{23}$ with the three terms given by
suitable permutations of
\begin{eqnarray} \label{e22}
  \phi_{23}
  =\tilde \phi_{23}(\vec r_{23})
  e^{i \vec k_1\vec r_1+i \vec K_{23}\vec R_{23}}
  \;,
\end{eqnarray}
where $\vec R_{23}=(m_2\vec r_2+m_3\vec r_3)/(m_2+m_3)$ is the center
of mass of the bound subsystem and $\vec K_{23}$ is the conjugate wave
vector.  A generalization of this three-body wave function is
\begin{eqnarray}
  \phi_{ij}=
  \tilde \phi_{ij}(\vec r_{ij})
  \exp\Big({i \sum_{k\ne i,j}\vec k_k\vec r_k+i \vec K_{ij}
    \vec R_{ij}} \Big)
  \;,\quad
  \Phi=\sum_{i<j}^N\phi_{ij}
  \;.
\end{eqnarray}
When all relative energies are small, $K_{ij}\simeq0$ and
$k_k\simeq0$, we obtain $\phi_{ij}\simeq \tilde \phi_{ij}(\vec r_{ij})$.

Generalization to an $N$-particle system was formulated by
Yakubovski{\u\i} who arranged the particles into all possible groups
of subsystems and thereby formally was able to include the correct
large-distance asymptotic behaviour for all cluster divisions
\cite{yak67}.  The decisive physical properties are related to the
division into clusters which for $N=3$ amounts to three possibilities.
The three Faddeev components are related to the number of divisions
and not the number of particles.  For $N > 3$ the number of cluster
divisions is much larger than $N$. For $N$ particles the wave function
is therefore written as a sum over possible clusters
\begin{eqnarray} \label{e24}
  \Phi
  =
  \sum\sub{clusters}\phi(\textrm{cluster})
  \;.
\end{eqnarray}
This method is often applied with success in nuclear physics
\cite{fad93,cie98,fil02b}.  In a dilute system two close-lying
particles are found much more frequently than any other cluster
configuration.  Then the dominating terms in the general cluster
expression in equation~(\ref{e24}) are the two-body clusters where the
remaining particles can be considered uncorrelated and described by
plane waves or in the mean-field approximation. The wave function then
reduces to the form
\begin{eqnarray} \label{e25}
  \Phi(\rho,\Omega)
  =
  \sum_{i<j}^N \phi_{ij}(\rho,\Omega)
  \;.
\end{eqnarray}

\subsubsection{Jastrow procedure.}

\label{sec:jastrow}

The Jastrow variational formulation was designed to account for
correlations \cite{bij40,din49,jas55}.  We will briefly comment on the
Jastrow ansatz, since it provides a physically transparent reason for
writing the wave function as a Faddeev sum in the dilute limit.  A
connection between the Jastrow ansatz for the relative wave function
\cite{jas55}
\begin{eqnarray}
  \Psi(\rho,\Omega) =
  \prod_{i<j}^N\psi(\vec r_{ij})
\end{eqnarray}
and the Faddeev formulation is possible. We write the two-body Jastrow
component as a mean-field term multiplied by a modification expected
to be important only at small separation, i.e., (omitting
normalization)
\begin{eqnarray}
  \psi(\vec r_{ij})
  =
  e^{-r_{ij}^2/(2Nb\sub t^2)}
  [1+\phi(\vec r_{ij})]
  \;,\qquad
  \phi(\vec r)=0 \quad\textrm{for}\quad r>r_0
  \;,
\end{eqnarray}
where we introduced the length scale $r_0$ beyond which deviations due
to correlations vanish. With equation~(\ref{e3}) this leads to the
relative wave function
\begin{eqnarray}
  \Psi(\rho,\Omega)
  =
  e^{-\rho^2/(2b\sub t^2)}
  \prod_{i<j}^N
  \Big[1 + \phi(\vec r_{ij})\Big]
  \\ \label{e29}
  =
  e^{-\rho^2/(2b\sub t^2)}
  \Big[ 1
  +\sum_{i<j}^N\phi(\vec r_{ij})+
  \sum_{i<j\ne k<l}^N\phi(\vec r_{ij})\phi(\vec r_{kl})
  +\cdots  \Big]
  \;.\quad
\end{eqnarray}
Through equation~(\ref{eq:Gauss_product}) the Gaussian mean-field
Hartree-ansatz is obtained for a non-interacting system in the
harmonic external field. For a homogeneous density distribution with
$b\sub t\to\infty$ the mean-field solution is a constant \cite{cow01}.
For a sufficiently dilute system it is unlikely that more than two
particles simultaneously are close in space, i.e., both $r_{ij}<r_0$
and $r_{kl}<r_0$. Therefore the expansion in equation~(\ref{e29}) can
be truncated after the first two terms, i.e.,
\begin{eqnarray}
  \prod_{i<j}^N\Big[1+
  \phi(\vec r_{ij})\Big]
  \simeq
  1+\sum_{i<j}^N\phi(\vec r_{ij})
  =
  \sum_{i<j}^N\bigg[
  \frac{1}{N(N-1)/2}+\phi(\vec r_{ij})
  \bigg]
  \;.
\end{eqnarray}
Redefining the two-body amplitude we end up with a Faddeev-like sum as
in equation~(\ref{e25}).

\subsection{Two-body $s$-wave correlations}

The conclusion from the preceding subsection is that a wave function
of the form
\begin{eqnarray}
  \Psi(\rho,\Omega)
  =
  F(\rho)\sum_{i<j}^N\phi(\rho,r_{ij})
\end{eqnarray}
incorporates both the mean-field properties through $F(\rho)$ and the
correlations in addition to the mean-field through the
Faddeev-components $\phi$.  We therefore decompose the angular wave
function $\Phi$ in equation~(\ref{eq:hyperspherical_wave_function})
(omitting indices $\nu$), into the symmetric expression of Faddeev
components $\phi$
\begin{eqnarray} \label{e31}
  \Phi(\rho,\Omega) 
  =
  \sum_{i<j}^N  \phi_{ij}(\rho,\Omega) 
  \approx
  \sum_{i<j}^N  \phi(\rho,r_{ij})
  \;.
\end{eqnarray}
where the last approximation assumes that only relative $s$-waves
contribute, leaving the dependence on the distance $r_{ij}$. Higher
partial waves could in principle be included but the numerical
complications would increase rather dramatically. Thus, we have
``only'' assumed relative $s$-waves between each pair of particles as
appropriate for small relative energies and large distances.  The
capability of this decomposition for large scattering length has been
demonstrated for $N=3$ by an application to the intricate Efimov
effect, which also arises precisely for small energies and large
distances \cite{fed93,jen97}.

The Faddeev ansatz in equation~(\ref{e25}) can be formally established
as a generalized partial wave expansion in terms of the hyperspherical
harmonic kinetic energy eigenfunctions.  The two-body $s$-wave
simplification then appears as a truncation of this expansion to
include only the lowest hyperharmonics for the description of the
remaining $N-2$ particles. Since this function is a constant we arrive
at equation~(\ref{e31}).  This $s$-wave assumption emphasizes two-body
correlations.  The method can be extended to include higher-order
correlations directly in the form of the wave function, e.g.,
three-body correlations, as suggested by \cite{bar99a}.

In conclusion, we believe that the Faddeev ansatz with two-body
amplitudes accounts for the important two-body correlations when the
system is sufficiently dilute, and at the same time keeps the
mean-field information about motion relative to the remaining
particles.  An extension of this technique would be a feasible, but
perhaps intricate, approach to study three-body correlations in denser
systems and in connection with the important process of three-body
recombination within $N$-body systems.

\section{General properties}

\label{sec:general-properties}

The method outlined above leads to the effective radial potential in
equation~(\ref{eq:radial.potential}) and the radial
equation~(\ref{eq:radial.equation}). This huge simplification is
hiding all the complications and the detailed information in the
angular eigenvalue computations.  The key quantity is then the
function $\lambda$, which determines the properties of the radial
potential.

The angular eigenvalue equation~(\ref{eq:angular_equation}) can by a
variational technique be rewritten as a second-order
integro-differential equation in the variable $\alpha_{N-1}$
\cite{sor01}.  For atomic condensates the interaction range is very
short compared with the spatial extension of the $N$-body system.
Using this short-range property of the interaction in the
integro-differential equation simplifies even further to contain at
most one-dimensional integrals.  The validity of our approximations
only relies on the small range of the potential, whereas the
scattering length can be as large as desired.

The general structure of the interaction between neutral atoms is an
attraction at longer distances arising from mutually induced
polarization.  At shorter distances the Pauli blocking dominates and
causes the effective interaction to be repulsive.  This is often
modelled by potentials similar to the van der Waals potential.  In the
present formulation it is possible to use any short-range potential
also with a finite repulsion at the core, e.g., a sum of two Gaussians
with repulsion at the origin and attraction at larger distances.
However, for the large distances crucial for the condensate's
properties, only the scattering length is important.  We therefore
first apply a Gaussian potential $V(r)=V_0\exp(-r^2/b^2)$ and study
dependence on the scattering length $a_s$ for a fixed range $b$.  It
is convenient to measure the strength of the interaction in units of
the Born-approximation $a\sub B$ of the scattering length
\begin{eqnarray} \label{e33}
  a\sub B
  \equiv
  \frac{m}{4\pi\hbar^2}
  \int d^3\vec r_{kl}\; V(\vec r_{kl})
  \;,
\end{eqnarray}
which for the Gaussian potential is $a\sub B = \sqrt\pi
mb^3V_0/(4\hbar^2)$.  Physical results when $\rho\gg\sqrt{N}b$ are
independent of the shape of the potential \cite{sor02b}.

The effective two-body interactions can vary enormously for different
systems and different experiments.  Depending on the strength of the
interaction the two particles may form a bound state of relatively
small radius (nm) compared to the typical size ($\mu$m) of a
Bose-Einstein condensate.  For the alkali atoms there are usually
several of such bound two-body states.  Scattering of two atoms at
sufficiently low relative energy depends only on the two-body $s$-wave
scattering length $a_s$.  Large distances appropriate for dilute
systems can then be expected to be determined almost entirely by
$a_s$.  At higher densities also the effective range of the
interaction may be significant.

We then first solve the angular equation by the method of finite
differences \cite{sor02b}.  The basis points are chosen to catch the
rapidly varying parts of the wave function and the finite short-range
potential.  This implies that the points vary strongly with
hyperradius and particle number. With the angular eigenvalue and wave
function we then continue to solve the much simpler radial equation
where only one adiabatic potential is included.

\subsection{Angular potential}

Two-body interactions are responsible for the properties of the
many-body system.  In our formulation, first the properties of the
angular eigenvalues are determined and next they enter decisively the
effective potentials and the radial equations.  Qualitatively the
results depend on the sign of the scattering length and the number of
two-body bound states.  This is understandable as the atoms in a
dilute system at low energy effectively interact as in a two-body
scattering situation.  Higher-order processes seldom occur and do not
contribute to the properties of the dilute system.

\begin{figure}[htb]
  \centering
  \input{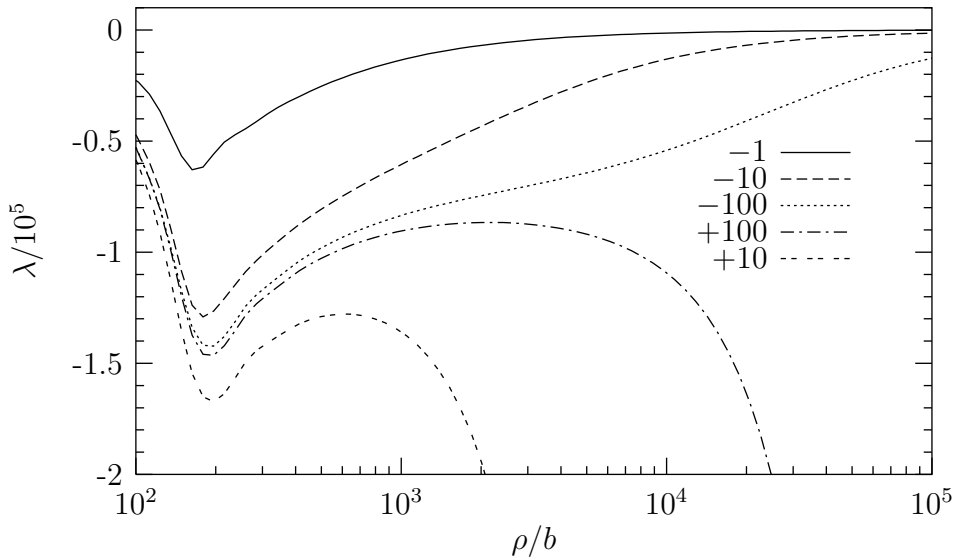}
  \caption[] {The lowest angular eigenvalues $\lambda$ for $N=100$
    bosons interacting via a Gaussian two-body potential
    $V(r)=V_0\exp(-r^2/b^2)$ with zero or one bound two-body states.
    The scattering lengths $a_s /b$ are indicated on the figure.}
  \label{fig:lambda_numerical}
\end{figure}

The study in \cite{sor03a} included variations of the interaction
strength and the number of particles.  We show in
figure~\ref{fig:lambda_numerical} the lowest angular potential from
equation~(\ref{eq:angular_equation}) for $N=100$ for Gaussian two-body
interactions with various scattering lengths.
When $a_s = -b$ (solid line) the potential has no bound two-body
states. The lowest angular eigenvalue is zero at $\rho=0$, decreases
then through a minimum as a function of $\rho$ and continues
afterwards to approach zero at large hyperradii as $a_s/\rho$.
Increasing the attraction (broken lines) decreases all angular
eigenvalues for all $\rho$-values. The details at smaller hyperradii
hardly change with large variations of the scattering length.
However, at larger distances the approach towards zero is converted
into a parabolic divergence as soon as the scattering length jumps
from negative (dotted) to positive (dot-dashed) corresponding to the
appearance of a bound two-body state.

Generally, an attractive finite-range interaction can support a
certain number $\mathcal N\sub B$ of two-body bound states for both
positive and negative scattering lengths. Then the lowest angular
eigenvalues, $\lambda_0, \lambda_1,\ldots,\lambda_{\mathcal N\subsub
  B-1}$, describe these bound two-body states within the many-body
system at large hyperradii, i.e., they diverge to $-\infty$ as seen in
figure~\ref{fig:lambda_numerical}. The next eigenvalue
$\lambda_{\mathcal N\subsub B}$ converges to zero at large distance
and corresponds to the first ``two-body-unbound'' mode.  Through the
derived adiabatic potential this mode is responsible for the
properties of atomic Bose-Einstein condensation, where no
clusterization is allowed.

The detailed numerical analysis in \cite{sor03a} resulted in a
parametrization for the behaviour of these $\lambda$-functions.  Here
we restrict ourselves to attractive two-body interactions in two
different regimes: i) no bound two-body states and $a_s<0$, and ii)
one bound two-body state and $a_s>0$. For hyperradii exceeding a lower
limit $\rho_0$, which roughly is at the minimum, i.e., $\rho > \rho_0
\equiv 0.87N^{1/2}(b/|a_s|)^{1/3}b$ the analytical expressions are
\cite{sor03a}
\begin{eqnarray}
  \lambda\sub{a}(N,\rho)
  &=&
  -|\lambda_\delta(N,\rho)|\Bigg(1+\frac{0.92N^{7/6}b}{\rho}\Bigg)
  \nonumber\\
  &&
  \times\left\{
    \begin{array}{ll}
      1-\exp\Big(-\frac{|\lambda_\infty(N)|}{|\lambda_\delta(N,\rho)|}\Big)
      & \textrm{when } a_s<0 \;, \\
      \frac{|\lambda_\infty(N)|}{|\lambda_\delta(N,\rho)|}+
      \frac{|\lambda^{(2)}(\rho)|}{|\lambda_\delta(N,\rho)|}
      & \textrm{when } a_s>0 \;,
    \end{array}
  \right.
  \label{eq:schematic}
\end{eqnarray}
where $\lambda_\delta$ is the expectation value of $\hat h_\Omega$ for
the zero-range interaction $V_\delta(\vec r) = 4\pi\hbar^2a_s
\delta(\vec r)/m$ in a constant angular wave function
$\Phi(\rho,\Omega)=$constant, i.e.,
\begin{eqnarray}
  \lambda_\delta(N,\rho)
  &=&
  \sqrt{\frac{2}{\pi}}
  \;
  \gammafktB{3N-3}{3N-6}
  \; N (N-1) \; \frac{a_s}{\rho}
  \;\stackrel{N\gg1}{\longrightarrow}\;
  \frac{3}{2}\sqrt{\frac{3}{\pi}}N^{7/2}\frac{a_s}{\rho}
  \label{eq:lambda_delta}
  \;,\\
  \label{eq:lambda_infty}
  \lambda_\infty(N)
  &=&
  -1.59N^{7/3}
  \;,
  \\
  \lambda^{(2)}(\rho)
  &=&
  E^{(2)}\frac{2m\rho^2}{\hbar^2}
  \;,\quad
  E^{(2)}=-\frac{\hbar^2}{m|a_s|^2}c
  \;.
\end{eqnarray}
The number $c$ approaches unity when the scattering length becomes
very large.  The factor $(1+0.92N^{7/6}b/\rho)$ reflects dependence on
potential details like the finite range $b$ of the Gaussian two-body
interaction.  At $\rho\sim N^{7/6}|a_s|$ we have $\lambda_\delta \sim
\lambda_\infty \sim \lambda^{(2)}$.  For small hyperradii $\rho <
\rho_0$ we use for all $a_s$ the perturbation result obtained as the
expectation value of the two-body interaction in a constant angular
wave function, i.e.,
\begin{eqnarray} \label{e36}
  \lambda\sub{a} 
  =
  \lambda^{(0)}(\rho)
  \equiv
  \frac{mV(0)N^2\rho^2}{\hbar^2}
  \;,\quad
  {\rm for}\;\rho<\rho_0
  \;,
\end{eqnarray}
where we use equation~(\ref{e36}) for small $\rho$ when
$|\lambda^{(0)}(\rho)|$ is smaller than the expression
equation~(\ref{eq:schematic}). Then the angular eigenvalue $\lambda$
is defined analytically for all $\rho$.  These expressions describe
accurately the results of full numerical computations for any two-body
interaction as soon as $\rho$ is larger than $\rho_0$.  We should
emphasize that the small-distance region where $\lambda\sub{a} =
\lambda^{(0)}(\rho)$ is sensitive to the specific choice of two-body
interaction.

The results of the parametrization in equations~(\ref{eq:schematic})
and ({\ref{e36}) are illustrated in figure~\ref{fig:schematic} for
  \mbox{$N=100$} for a larger range of scattering lengths than in
  figure~\ref{fig:lambda_numerical} in order to show the quality of
  the parametrization.
\begin{figure}[htb]
  \centering
  \input{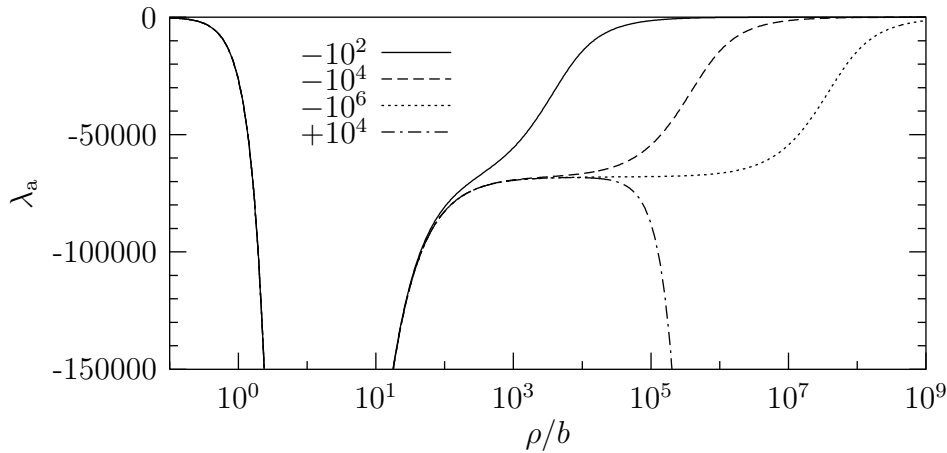}
  \caption[]
  {The angular eigenvalue $\lambda$, equations~(\ref{eq:schematic})
    and (\ref{e36}), for $N=100$ as function of $\rho$ for the
    different scattering lengths given on the figure in units of the
    range $a_s/b$.}  \label{fig:schematic}
\end{figure}
The pronounced deep minimum at $\rho\sim\rho_0$ is in the region
depending on the two-body potential and reflects only the qualitative
behaviour of the numerically correct lowest angular eigenvalue. After
this strongly attractive region at small $\rho$ the eigenvalues
approach zero. As the size of the scattering length increases the
eigenvalue develops a plateau at a constant value $\lambda_\infty$
independent of $a_s$. Eventually at large $\rho$ the eigenvalues
vanish as $\lambda_\delta$ for $a_s<0$ and diverges to $-\infty$
when $a_s>0$.

When $a_s<0$ the analytic and the correct eigenvalues both exceed
the asymptotic zero-range result, i.e., $\lambda\sub{a} \ge
\lambda_\delta$ for all hyperradii.  This means that the true ground
state energy is higher than the energy obtained with the zero-range
interaction.  Thus the ground state energy from our model is higher
than the mean-field energy. The origin of this sequence of energies is
that the zero-range interaction inevitably leads to diverging energies
for smaller distances. The present model avoids this non-physical
short-range collapse.

When $a_s>0$ the interaction is effectively repulsive at large
hyperradii and we find analogously that an analytical expression in
this case for the second angular eigenvalue obeys $\lambda\sub{a} \le
\lambda_\delta$ for all hyperradii, due to the divergence of
$\lambda_\delta\to+\infty$ as $\rho\to0$.  Correspondingly we get
energies smaller than the zero-range mean-field result in the positive
$a_s$-case.  \etalcite{Bohn}{boh98} obtained in this case only
energies higher than the Gross-Pitaevskii energy.

\subsection{Radial potential}

The parametrization in equation~(\ref{eq:schematic}) leads to an
analytic expression for the radial potential.  We can then also study
the properties of the radial potential and derive physical quantities
like the energy and the root-mean-square separation between bosons.
In particular the attractive two-body potentials generally give rise
to a large number of negative-energy many-body states.  Using the
method described in \cite{khu02} it is possible to estimate the number
$\mathcal N$ of such bound states, i.e.,
\begin{eqnarray} \label{e39}
  \mathcal N
  \simeq
  \frac{\sqrt{2m}}{\pi\hbar}
  \int \sqrt{|U^{(-)}(\rho)|} {\rm d} \rho 
  \;,
\end{eqnarray}
where $U^{(-)(\rho)}$ denote the negative part of the radial potential
$U(\rho)$.

\subsubsection{Features of the analytic expression.}

The radial potential obtained from equation~(\ref{eq:schematic}) is
shown in figure~\ref{fig:simul} as function of the hyperradius for a
series of different particle numbers and scattering lengths.  The
strongly-varying short-distance dependence is omitted to allow focus
on intermediate and large hyperradii. When an intermediate barrier is
present the condensate is described as the state of lowest energy
located in the minimum at large hyperradius. This minimium exists for
$a_s<0$ when $N|a_s|/b\sub t < 0.5$ as established in
\cite{boh98,sor02b}

The behaviour at very small hyperradii can be constructed from
equation~(\ref{e36}). However, now the central value of the two-body
interaction enters explicitly and the resulting radial potential
therefore depends on the short-distance behaviour of this interaction.
This model-dependence extends to larger distances where the
perturbation expression in equation~(\ref{e36}) is invalid.  This
region of $\rho$ up to $\sqrt{N}b$ is not very interesting in the
present context and we therefore only crudely connected the analytic
parametrization in equation~(\ref{eq:schematic}) and the expression in
equation~(\ref{e36}) to allow extraction of the model-independent
result.

Moving alphabetically in figure~\ref{fig:simul} from (a)-(f) we first
in (a)-(d) maintain the particle number $N=6000$ while only the
scattering length $a_s$ varies. From (d)-(f) we maintain
$a_s/b = -0.35$ and vary $N$.
\begin{figure}[htb]
  \centering
  \input{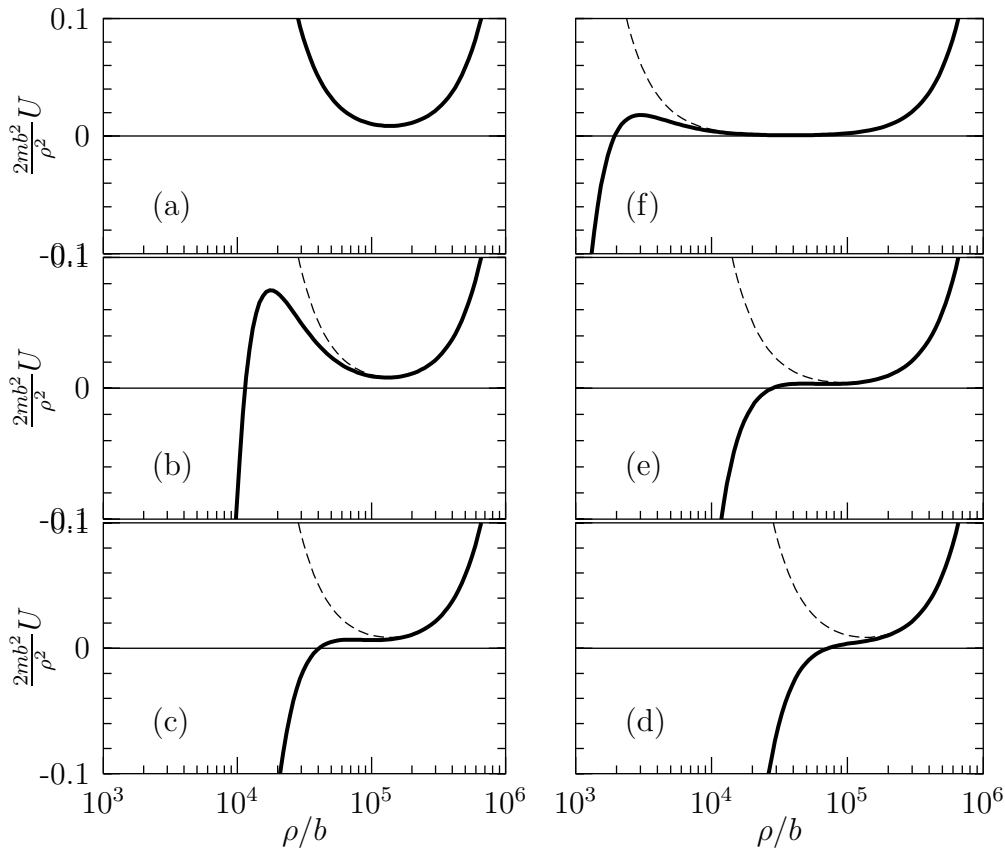}
  \caption[Radial potentials as a function of $N$ and $|a_s$]
  {Radial potentials with $b\sub{t}/b=1442$ and (a) $N=6000$,
    $a_s\sim0$; (b) $N=6000$, $a_s/b=-0.05$; (c) $N=6000$,
    $a_s/b=-0.18$; (d) $N=6000$, $a_s/b=-0.35$; (e) $N=3000$,
    $a_s/b=-0.35$; (f) $N=500$, $a_s/b=-0.35$.  The dashed lines are
    with $a_s=0$.  The divergence $U(\rho)\to +\infty$ when $\rho\to0$
    is not shown.}  \label{fig:simul}
\end{figure}
In (a) the two-body interaction is zero, $a_s=0$, leading to a
vanishing angular eigenvalue, $\lambda=0$. The effective radial
potential then consists only of centrifugal barrier and external field
with one minimum. In (b) we turn on an attractive potential,
$a_s=-0.05b$, sufficiently strong to overcompensate for the
centrifugal repulsion and create a second minimum in the radial
potential at smaller hyperradius.  An intermediate barrier is left
between the two minima at small and large hyperradii.  A further
increase of the attraction in (c) removes the barrier while leaving a
smaller, flat region. The negative-potential region around the minimum
at small hyperradius is now even more pronounced.  This tendency is
continued in (d) with a stronger attraction.

With the scattering length from (d), $a_s=-0.35b$, and a
decreasing number of particles the intermediate barrier is slowly
restored while moving to smaller hyperradii. In (e) for $N=3000$ a
barrier is about to occur, and in (f) with only $N=500$ an
intermediate barrier is again present between a minimum at small and
large hyperradii.

\subsubsection{Large scattering length}

\label{sec:large-scatt-length}

A large scattering length implies through
equation~(\ref{eq:schematic}) an intermediate region in hyperradius
where the angular potential is almost constant.  More specifically,
when $\rho<N^{7/6}|a_s|$ and $\lambda \simeq \lambda_\infty$ two of
the terms in the radial equation~(\ref{eq:radial.potential}) add to a
negative value
\begin{eqnarray} \label{e57}
  \frac{\lambda_\infty}{\rho^2}+ \frac{(3N-4)(3N-6)}{4\rho^2} <0 \;,
\end{eqnarray}
which implies that no repulsive barrier is present. Then the effective
potential is $\rho^{-2}$ until the trap begins to dominate.

We show in figure~\ref{fig:u_schematic} the analytic radial potential
corresponding to one of the angular eigenvalues from
figure~\ref{fig:schematic}.  We observe that the radial potential is
negative in a large range of hyperradii which can be divided into
three different regions.  For small hyperradii the radial potential
has a minimum.  For intermediate hyperradii the angular potential is a
constant and therefore the radial potential behaves as $-1/\rho^2$.
This is from figure~\ref{fig:schematic} seen to appear for $\rho/b$
between $10^2$ and $10^5$.  When $\rho\ge N^{7/6}|a_s|$ the angular
potential vanishes as $-1/\rho$, so the radial potential vanishes as
$-1/\rho^3$, although not clear on the figure.  Finally the trap
$\propto\rho^2$ dominates with positive contributions at large
hyperradii.

\begin{figure}[thb]
  \centering
  \input{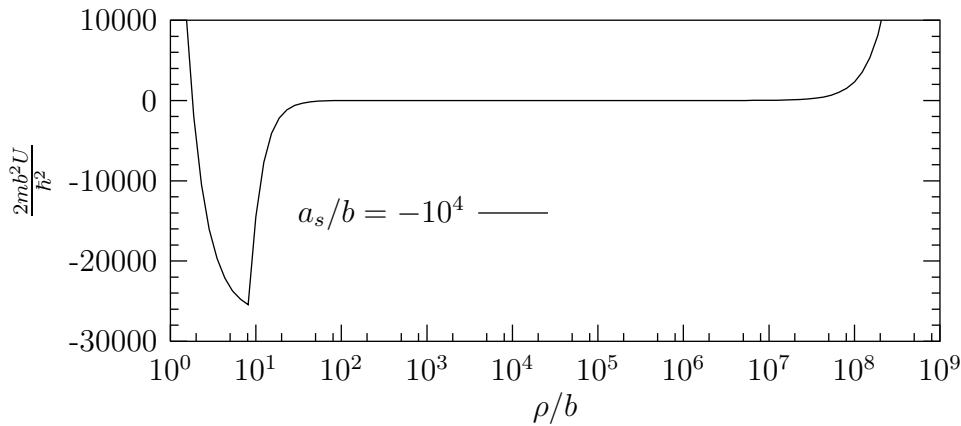}
  \caption  []
  {Analytic radial potential obtained from equations
    (\ref{eq:radial.potential}) and (\ref{eq:schematic}) for $N=100$
    and $a_s/b=-10^4$.}
  \label{fig:u_schematic}
\end{figure}

The bound states in this potential can be divided into groups
according to their hyperradial extension. The total number of such
states is written as $\mathcal N = \mathcal N_1 + \mathcal
N_E+\mathcal N_2$ where $\mathcal N_1$, $\mathcal N\sub{E}$, and
$\mathcal N_2$ are the number of states located respectively in the
attractive pocket at small hyperradii, in the intermediate $-1/\rho^2$
region and at hyperradii large compared with the scattering length.

With the analytic expressions inserted into equation~(\ref{e39}) to
obtain the effective potential we find a crude estimate of the number
of self-bound states in the pocket to be $\mathcal N_1\simeq 1.3
N^{3/2}$. This number depends on the properties of the two-body
potential but the $N^{3/2}$ scaling remains unchanged for all
short-range interactions. The outer region also supports bound states
when the trap length $b\sub{t}$ is sufficiently large, i.e., $b\sub t
\gg N|a_s|$, where we analogously find that $\mathcal N_2 \simeq 0.78
N^{7/6}$. This number may be severely influenced by the confinement of
the external trap, but again the $N^{7/6}$ scaling remains unchanged.

The intermediate region is only present when the scattering length is
relatively large.  This region exist when 
\begin{eqnarray}
  b \ll \frac{\rho}{N^{7/6}} \ll |a_s|
  \;,
  \label{eq:intermediate_region}
\end{eqnarray}
where these limits correspond to values of the hyperradius larger than
$\rho\sub{min} = N^{7/6}b$ and smaller than $\rho\sub{max} =
N^{7/6}|a_s|$.  The number of states $\mathcal N\sub{E}$ located in
this region is again obtained from equations~(\ref{e39}), (\ref{e57})
and (\ref{eq:radial.potential}), see
\cite{nie01,sor02}. This gives
\begin{eqnarray} \label{e42}
  \mathcal N\sub{E} &\simeq& \frac{|\xi|}{\pi}
  \ln\bigg(\frac{\rho\sub{max}}{\rho\sub{min}}\bigg) \;,\\
  \xi^2 &\equiv& -\lambda_\infty-\frac{(3N-4)(3N-6)}{4}-\frac{1}{4} \approx
  1.59N^{7/3}  \; ,
\end{eqnarray}
where we used equation~(\ref{eq:lambda_infty}) and assume $N$ is
large.  The number of these bound states is then
\begin{eqnarray}
  \mathcal N\sub{E}
  &\simeq&
  0.40N^{7/6}
  \ln\Bigg(\frac{|a_s|}{b}\Bigg)
  \;.
\end{eqnarray}
They are located in the region, where the radial potential behaves as
$1/\rho^2$, which is the generic form of the potentials giving rise to
the Efimov states in three-body systems \cite{fed93,nie01}.  These
states have characteristic scaling properties relating neighboring
values of energies and mean square radii.  The number of states
depends logarithmically on the size of the $1/\rho^2$-region as in
equation~(\ref{e42}), analogous to the three-body Efimov states.
These states were therefore denoted many-body Efimov states
\cite{sor02}.

This estimate assumed that the external trap has no influence on the
hyperradial potential for $\rho <\rho_{max}$. However, when the trap
length $b\sub{t}$ is sufficiently small, i.e., when $\rho\sub{trap} =
\sqrt{N}b\sub{t} < N^{7/6}|a_s|$, the extension of the plateau is
truncated at large hyperradii.  The number of states can then be
estimated by substituting $\rho\sub{max}$ with $\rho\sub{trap}$ in
equation~(\ref{e42}).  This yields
\begin{eqnarray}
  \mathcal N\sub{E}
  &\simeq&
  0.40N^{7/6}
  \ln\Bigg(\frac{\sqrt{3/2}b\sub t}{N^{2/3}b}\Bigg)
  \;.
\end{eqnarray}

The plateau can not exist when the external potential dominates
already at small distances, i.e., for short trap lengths when
$\rho\sub{trap}<N^{7/6}b$ or equivalently $N\gtrsim N\sub{max} \equiv
(b\sub t/b)^{3/2}$.  This maximum number of particles allowing a
plateau and the resulting Efimov-like states is for a realistic ratio
of $b\sub t/b=1442$ therefore obtained to be $N\sub{max}\simeq55000$.
This estimate is rather uncertain but it illustrates that too many
particles not only exclude stability of the condensate but also the
existence of the spatially extended Efimov-like structures which
otherwise might play a role in the recombination processes.

The number of Efimov-like states $\mathcal N\sub E$ increases strongly
with $N$ as seen in table~\ref{table:NE} for $b\sub t/b=1442$ for a
few particle numbers. These estimates are more precise than in
previous work where we obtained $\mathcal N\sub E=28$ for $N=20$ for
the same parameters \cite{sor02}.
\begin{table}[htb]
  \centering 
  \begin{tabular}{|c||c|c|c|c|c|} 
    \hline 
    $N$ & $10$ & $20$ & $100$ &  $1000$ & $10000$ \\ 
    \hline\hline
    $\mathcal N\sub{E}$ & 35 & 72 & 380 &  3632 & 24810 \\
    \hline 
  \end{tabular} 
  \caption[] {The number $\mathcal N\sub{E}$ of Efimov-like states for 
    $b\sub t/b=1442$ and $|a_s|\to\infty$.}  
  \label{table:NE}
\end{table}

The energies and mean square radii of the Efimov-like states are related
by the expressions
\cite{nie01}
\begin{eqnarray}
  E_n
  =
  -\frac{\hbar^2}{2m\langle\rho^2\rangle_n}\frac23(1+\xi^2)
  \;,\quad
  E_n=E_0e^{-2\pi n/\xi}
  \;,
  \label{eq:efimov_energy}
\end{eqnarray}
where the exponential dependence on both the strength $\xi$ of the
effective potential and the number of excited states is highlighted.

Let us assume that the trap length is large and not responsible for
terminating the plateau at large distance.  We can then crudely assume
that the mean square hyperradii of the first and last Efimov-like
states are given by $\rho\sub{min}^2 = N^{7/3}b^2$ and
$\rho\sub{max}^2 = N^{7/3}|a_s|^2$, respectively. Using
equation~(\ref{eq:efimov_energy}) we then obtain the energies of the
first and last Efimov-like states
\begin{eqnarray} \label{e46}
  E\sub{first} \simeq -\frac{\hbar^2}{2mb^2}
  \;,\quad
  E\sub{last} \simeq -\frac{\hbar^2}{2m|a_s|^2}
  \;.
\end{eqnarray}
which turn out to be independent of the particle number $N$.  These
energies remind of the kinetic energy scale of strongly bound two-body
states and the expression for a weakly bound or a resonance two-body
energy, respectively. This does not mean that the average distance
$\bar r$ between two particles in these many-body states also are
given by $b$ and $a_s$. In fact, $\bar r$ contains an additional
$N$-dependent factor, i.e., $\bar r \approx N^{2/3}b, N^{2/3}|a_s| $
for the two cases.  These constant energy limits imply that the
density of Efimov-like states increases with particle number precisely
as the interval scales, i.e., $N^{7/6}$.

\subsubsection{Decay and collapse.}

The Bose-Einstein condensate is intrinsically unstable and decays
spontaneously, e.g., into lower lying dimer states.  Recombination of
two particles into a lower-lying (bound) state is possible by emission
of a photon, but the rate is strongly enhanced when a third particle
is involved instead of the photon.  This three-body recombination
process has been suggested to be important in Bose condensates
\cite{adh01b,ued03}.  The outcome of dimers can not be distinguished
directly from these very similar processes.  Molecular formation from
two cold atoms, enhanced by tuning the Feshbach resonance, corresponds
to absorption of a photon and creation of a meta-stable structure.
The related change of the surrounding medium could lead to instability
collectively involving many particles, and much faster decays better
described as a collapse \cite{adh02b}.

In the experiments by \etalcite{Donley}{don01} a condensate was first
created with effectively zero interaction, i.e.,~zero scattering
length as in figure~\ref{fig:simul}a.  The radial wave function is
then located at relatively large distances in the minimum created by
the compromise between centrifugal barrier and external field.  The
attractive pocket at small distances is not present and the condensate
appears as the ground state in this potential.  Both the radial
potential and the wave function are shown in
figure~\ref{fig:colsequence}.  The effective interaction was then
suddenly changed by tuning a Feshbach resonance \cite{cor00} to obtain
a large and negative scattering length \cite{don01}.  The measurement
showed a burst and a remnant of coherent atoms. This was interpreted
and explained as formation of dimers via the two-body resonance, a
burst of dissociating dimers and a remnant of an oscillating mixture
of coherent atoms and coherent molecules \cite{kok02b,mac02,koh03b}.

In our formulation the effective potential is suddenly altered by a
change of the underlying two-body interaction. The corresponding new
radial potential, shown in figure~\ref{fig:colsequence}, has a
pronounced attractive region able to support a number of bound states.
The initial wave function is no longer a stationary state in the new
potential and a motion is started towards smaller hyperradii, where it
would be reflected from the wall of the centrifugal barrier.  The
system would then oscillate between the centrifugal barrier and the
wall of the external field.

This makes the unrealistic assumption that no other degrees of freedom
are exploited, e.g., the angular dependence of the wave function or
molecular bound states described by other adiabatic potentials.  Thus
direct population of two-body bound states and resonances are not
allowed.  This requires in addition inclusion of the adiabatic
potential asymptotically describing these states.  This is entirely
possible within our model, but it constitutes a major new numerical
investigation where coherent atoms and molecules, oscillations between
them, and (three-body) recombination are studied in the same
framework.

In the present work we confine ourselves to the scenario of
macroscopic contraction, where the density rapidly increases and
dimers quickly are produced and subsequently ejected from the trap. We
shall in the following make qualitative estimates of the three-body
recombination rate producing the dimers. Whether this process is
significant or not remains to be seen.

To study the process we maintain the chosen degrees of freedom
described by one adiabatic potential. We expand the initial wave
function on the eigenfunctions in the new adiabatic potential.  The
dominating states in this expansion are the highest-lying Efimov-like
states now present because of the large scattering length which
produces the plateau region and the $\rho^{-2}$ potential.  These
states have a similar large spatial extension as the initial wave
function.  The resulting non-stationary wave function provides a
specific oscillation time.  After a quarter of a period the extension
of the system has reached its minimum.  The wave function at this time
$T$ is also shown in figure~\ref{fig:colsequence}.

\begin{figure}[thb]
  \centering
  \input{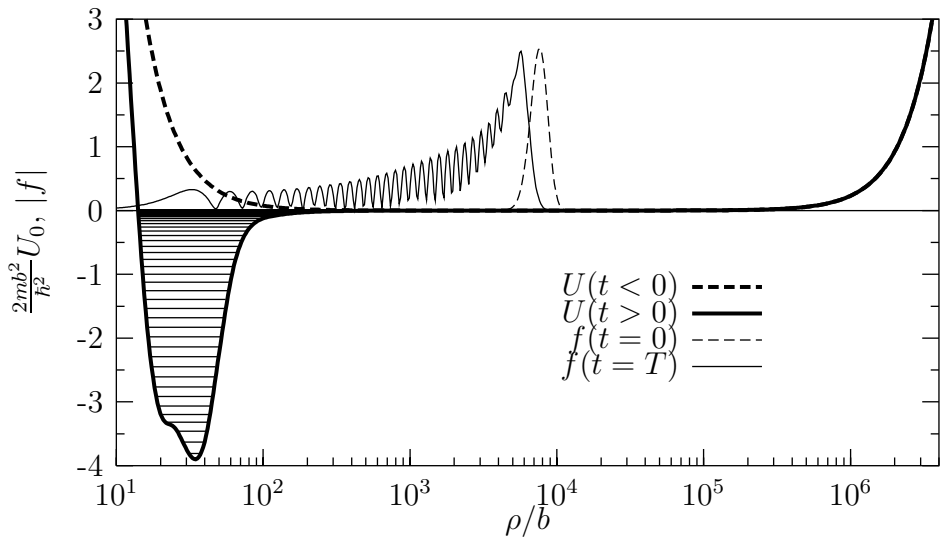}
  \caption[] {Wave functions $f$ and effective hyperradial potentials
    $U$ in dimensionless units as function of hyperradius for $N=20$.
    The scattering length is zero up to the time $t=0$ and then
    suddenly changes to be large and negative at later times $t>0$.
    Potentials and the corresponding wave functions are sketched for
    $t=0$ and at a time T after a quarter of a period. The horizontal
    lines show the stationary negative-energy states for $t>0$.}
  \label{fig:colsequence}
\end{figure}

The recombination probability increases with decreasing hyperradius
due to the higher density, i.e., several particles are close in space
and therefore much more likely recombine into molecular states.  The
time scale for three-body recombination is given by
$N(t)=N(0)\exp(-t/T\sub{rec})$, where $N$ is the number of atoms in
the condensate.  This is as a function of the average hyperradius
$\bar\rho$ estimated by \cite{sor03a}
\begin{eqnarray}
  T\sub{rec}
  =
  \frac{2m\bar\rho^6}{\hbar|a_s|^4N^3} \;.
  \label{eq:threebodyrectime}
\end{eqnarray}
This recombination time for the highest-lying Efimov-like states
($\bar\rho \approx N^{7/3} |a_s|$) can then be compared to the time
scale for motion in the condensate which is given by $T\sub{trap}
\approx 2\pi/\omega$.  We find
\begin{eqnarray}  
  \label{e47}
  \frac{T\sub{rec}}{T\sub{trap}}
  \approx
  \frac{N^2}{\pi}
  \Bigg(\frac{N|a_s|}{b\sub t}\Bigg)^2
  \;.
\end{eqnarray}
Thus, close to the limit of stability established as $N|a_s|/b\sub t
\sim 0.5$, we have $T\sub{rec} \gg T\sub{trap}$ ($N \gg 1$) and the
recombination process is rather slow for these highest-lying
Efimov-like states. Still the lifetime must in all cases be shorter
than for the initially created condensate because the density is
larger.

If these Efimov-like states are populated in experiments where the
potential suddenly is changed from figure~\ref{fig:simul}a to
figure~\ref{fig:simul}d,e they could possibly be indirectly observed.
A signature of this many-body Efimov effect would be observation of
the diatomic molecules formed in the recombination process and with
the estimated rate $T\sub{rec}$ from equation~(\ref{e47}). The rate
should then be inversely proportional to the square of the scattering
length reached after changing the potential.  The dimers themselves
can probably not be distinguished from this and other processes, but
the measured rate of dimers can possibly be separated into different
characteristic components.

These Efimov-like states may exist as quasistationary states
essentially decoupled from all the other many-body degrees of freedom.
The recombination time or the corresponding width $\Gamma\sub{Efimov}
= \hbar/ T\sub{rec}$ of the Efimov-like states indicates the degree of
decoupling. Using equations~(\ref{eq:efimov_energy}) and (\ref{e46})
we obtain
\begin{eqnarray}
  \frac{\Gamma\sub{Efimov}}{E_n -E_{n-1}}
  \approx
  \frac{1}{4\pi^2N^{11/6}}
  \Bigg(\frac{b}{a_s}\Bigg)^2
  \; , \quad
  \frac{\Gamma\sub{Efimov}}{E_n -E_{n-1}}
  \approx
  \frac{1}{4\pi^2N^{11/6}}
\end{eqnarray}
for the first and last Efimov-like states, respectively.  The
couplings compared to the level spacings are small and decreasing with
$N$.  Thus the identities of these states could be very well preserved
within the many-body system.  Still their lifetimes due to
recombination processes can be very large compared to the time scale
defined by the external field.  These negative-energy self-bound
many-body states should essentially maintain their spatial extension
after the external field is switched off. This is in clear contrast to
positive energy states where only the ion trap prevents expansion.
Thus a relatively slow time evolution of the density distribution
without external field should be characteristic for these very weakly
coupled many-body Efimov states.

\section{Connections to the mean-field approximation}

\label{sec:conn-mean-field}

The mean-field is often used to describe a condensate.  A Hartree
product of single-particle wave functions describes successfully a
Bose-Einstein condensate of a dilute, weakly interacting gas of
pointlike particles with $n|a_s|^3\ll1$, where $n$ is the density.
The mean-field validity condition is then fulfilled, i.e., the mean
free path is long compared to the interaction range of the system
defined by the scattering length.  The low-energy scattering
properties expressed by the scattering length are then clearly
decisive.  In the following we first comment on the choice of
interaction and second on the differences between the mean-field
method and the hyperspherical adiabatic method.  Finally we discuss
the conditions of validity.

\subsection{The two-body interaction}

The choice of the interactions should be consistent with the Hilbert
spaces for the different methods.  In the mean-field treatment a
zero-range interaction is often applied
\begin{eqnarray}
  V_\delta(\vec r)
  =
  \frac{4\pi\hbar^2a_s}{m}
  \delta(\vec r)
  \;,
  \label{eq:meanfield_potential}
\end{eqnarray}
where $a_s$ is the two-body $s$-wave scattering length. This limit can
be obtained from a finite-range potential where the range approaches
zero and the strength is appropriately adjusted.  We use a
finite-range Gaussian interaction
\begin{eqnarray}
  V\sub G(\vec r)=V_0e^{-r^2/b^2}
  \;,
  \label{eq:Gauss-pot}
\end{eqnarray}
where the Born-approximation $ a\sub B$ to the scattering length then
is a measure of the strength, see equation~(\ref{e33}).  The Gaussian
is in the limit when $b\to0$ a representation $\Delta_b(\vec r)$ of
the Dirac $\delta$-function
\begin{eqnarray}
  \Delta_b(\vec r)
  \equiv\frac{1}{\pi^{3/2}b^3}e^{-r^2/b^2}
  \;,\quad
  1=\int d^3\vec r
  \;\Delta_b(\vec r)
  \;.
\end{eqnarray}
We rewrite equation~(\ref{eq:Gauss-pot}) as
\begin{eqnarray} \label{e50}
  V\sub G(\vec r)=\pi^{3/2}b^3V_0\Delta_b(\vec r)
  =
  \frac{4\pi\hbar^2a\sub B}{m}\Delta_b(\vec r)
  \;,
\end{eqnarray}
which has the same form as equation~(\ref{eq:meanfield_potential}),
but with $a\sub B$ instead of $a_s$.  Then for $a_s=a\sub B$ we have
\begin{eqnarray}
  \lim_{b\to0}V\sub G(\vec r)=V_\delta(\vec r)
  \;.
\end{eqnarray}
However, $a_s = a\sub B$ is only valid when $|a\sub B|/b \to 0$, which
is rarely the case.

The limit of vanishing range $b$ can be reached in several ways, e.g.,
as in equation~(\ref{e50}) with a constant $a\sub{B}$ or with an
adjustment of $V_0$ to keep a constant $a_s$.  These limits differ
enormously and the optimum choice depends on the purpose and the
Hilbert space restricting the wave function.  If the low-energy
scattering properties are crucial the constant $a_s$ seems to be the
choice.  However, this does not lead to
equation~(\ref{eq:meanfield_potential}), but the strength of the
interaction should instead approach zero linearly with $b$.  In fact,
the scattering length is not even defined for the interaction in
equation~(\ref{eq:meanfield_potential}).  Still, the aim of computing
reliable energies in the mean-field approximation can be achieved with
this strength for dilute systems \cite{esr99b}.  The interaction and
the Hilbert space must be consistent, i.e., a renormalized interaction
follows a restricted space to produce the correct energy.  In this
case the Hilbert space is restricted to the mean-field product wave
functions.  Any extension to include features outside this restricted
space, for example two-body cluster structures, would be disastrous
\cite{fed01}. In other words the wave functions are very difficult to
improve even in perturbation theory.  Maintaining the finite-range
interaction with the correct scattering length then results in
different properties of the interaction even when the range approaches
zero on any scale defined by the physics of the problem. Thus the
mean-field product wave function with a realistic two-body potential
would also lead to disastrous results.

Clearly, the full Hilbert space with the correct interaction must
produce correct results. Whether the realistic interaction combined
with our choice of the space including two-body correlation amplitudes
can reproduce the main features is not apriori obvious. However, the
investigations summarized in the previous section demonstrate that the
energy of the mean-field approximation for dilute systems is
reproduced and the correct large-distance behaviour is at least
approximately obtained.  This asymptotic behaviour is determined by
the scattering length which only implicitly is contained in a given
combination of range and strength of the Gaussian interaction.  This
implies that our Hilbert space must account properly for the crucial
correlations necessary for an accurate description at large distances.

\subsection{Hyperspherical formulation with the zero-range interaction}

A reformulation of the mean-field in hyperspherical coordinates was
given by \etalcite{Bohn}{boh98}.  They assumed an angular wave
function, where all correlations are neglected, and a
$\delta$-interaction, equation~(\ref{eq:meanfield_potential}), is used
precisely as in the mean-field approximation.  This results in an
angular potential produced by the angular eigenvalue $\lambda_\delta$
in equation~(\ref{eq:lambda_delta}).  With this hyperspherical
potential they solve the radial equations.

Roughly speaking, our angular potential arising from the Gaussian
interaction is above $\lambda_\delta$ when $a_s<0$, and below when
$a_s>0$.  The ``exaggeration'' in \cite{boh98} of the zero-range
interaction is a result of including the $a_s/\rho$-divergence of
$\lambda_\delta$ also for small distances.  When $\rho$ approaches
zero or the scattering length diverges, these and other mean-field
methods yield disastrous results.

The mean-field interaction energy can be estimated as the expectation
value of the $\delta$-function interaction in
equation~(\ref{eq:meanfield_potential}) with a Hartree wave function
of Gaussian single-particle factors:
\begin{eqnarray}
  \psi(\vec r_i)
  =
  \frac{1}{\pi^{3/4}b\sub t^{3/2}}e^{-r_i^2/(2b\sub t^2)}
  \;,\\
  E\sub{int}
  =
  \frac{N(N-1)}2
  \int d^3\vec r_1 \; \psi^*(\vec r_1) V_\delta(r_1) \psi(\vec r_1)
  =
  \frac{2N(N-1)\hbar^2a_s}{\sqrt{\pi}mb\sub t^3}
  \;,
\end{eqnarray}
where we used $b\sub{t}$ as the size parameter for the wave function since
the confinement is due to the trap. This wave function is then the
lowest harmonic oscillator solution obtained without any two-body
interaction.

With hyperspherical coordinates this interaction energy is then related to the
angular eigenvalue:
\begin{eqnarray}
  E\sub{int}  = 
  \int_0^\infty d\rho
  f^*(\rho)
  \frac{\hbar^2\lambda_\delta(\rho)}{2m\rho^2}
  f(\rho)
  \;,
\end{eqnarray}
where $f$ is the normalized radial Gaussian function corresponding to
the Hartree form
\begin{eqnarray}
  f(\rho)
  =
  \sqrt{\frac{2}{\gammafkt{3N-3}b\sub t^{3N-3}}}
  \;
  \rho^{(3N-4)/2}
  \;
  e^{-\rho^2/(2b\sub t^2)} \; .
\end{eqnarray}
This radial wave function is not the correct solution obtained by
using the effective potential corresponding to $\lambda_\delta$.
However, only this Gaussian approximation allows an analytic
comparison between the hyperspherical and cartesian mean-field wave
functions.

\subsection{Properties of the wave functions}

The Hartree wave function is closely related to the hyperradial
function in the dilute limit and the Jastrow correlated wave function
is closely related to the Faddeev-like decomposition of the wave
function.  A direct comparison of the wave functions is not possible
in general as this requires an expansion on a complete set of basis
functions in one of the coordinate systems. The necessary calculations
involve non-reducible $3N$-dimensional integrals.

Instead we use the indirect relations provided in
section~\ref{sec:mean-field-descr}, where energy and average distance
between particles are characteristic features of the wave function.
For a given scattering length the energy $E$ is numerically obtained
for a Bose-Einstein condensate as a function of the particle number.
We then calculate the interaction energy defined as $E-E_0$ where $E_0
= 3N\hbar\omega/2$ is the energy of the non-interacting, trapped gas.
The results are shown in figure~\ref{fig:energy_N}.  For attractive
potentials the mean-field has a local minimum at large average
distance and much lower (diverging for zero-range) energies at small
average distances. The mean-field (quasistable) solution is located in
the minimum at large average distance. This minimum becomes unstable
for sufficiently large particle numbers. In the example of
figure~\ref{fig:energy_N}a no stable mean-field solution exists for
$N>1000$.  This is consistent with the stability criterion of about
$N|a_s|/b\sub{t}< 0.55$ as seen from the $x$-axis exhibited at the top
of the figure.

\begin{figure}[htb]
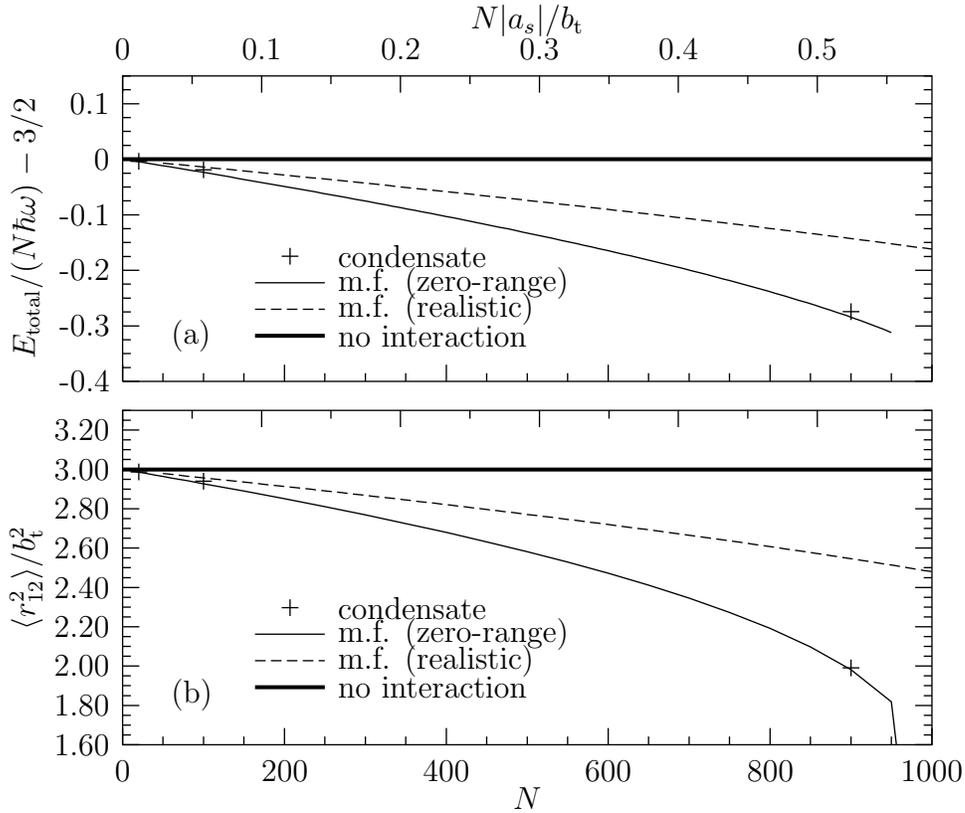

  \centering
  \input{fig6a.tex}
  \input{fig6b.tex}
  \caption[]
  {a) Gross-Pitaevskii energy as a function of $N$ for $a_s/b=-0.84$
    and $b\sub t/b=1442$.  Also shown are the hyperspherical
    calculation for three particle numbers and $a_s/b=-0.84$ ($a\sub
    B/b=-0.5$).  The dashed line shows the Gross-Pitaevskii energy for
    $a_s/b=-0.5$.  The $N|a_s|/b\sub t$-axis above only applies to
    $a_s/b=-0.84$.\\ b) Mean square distance between the particles for
    the cases of a).}
  \label{fig:energy_N}
\end{figure}

In the same figure we compare to results obtained with the present
method for three different particle numbers.  The interaction energies
are remarkably similar to those of the stable mean-field solution
where the scattering length in the Born approximation equals the
correct value.  We also show the results of the less attractive
zero-range interaction where the scattering length in the Born
approximation is the same as for the finite-range potential.  Now the
mean-field interaction energy is much less negative.  We should
emphasize that this comparison does not include the negative-energy
states supported by the attractive pocket at short distance, see
figure~\ref{fig:colsequence}.  They would appear below the
``condensate-like'' state shown in figure~\ref{fig:energy_N}a.

Using equations~(\ref{e3}) and (\ref{e21}) we compare in
figure~\ref{fig:energy_N}b $\langle r_{12}^2\rangle$ for the solutions
of the mean-field approximation and the hyperspherical methods.  The
mean square distance decreases with increasing particle number for all
calculations with an attractive potential. As $N$ approaches $1000$
the Gross-Pitaevskii mean-field radius approaches zero due to the
unavoidable collapse.  The same behaviour is seen for radii and
interaction energies, i.e., the average distance between particles
decreases until the condensate collapses and the size vanishes in the
mean-field while many-body bound states with smaller extension play a
role in the hyperspherical description.  Then also higher-order
correlations can be expected to be essential and result in
recombination processes.

The average distance is related to the interaction energy $E-E_0$.  In
a harmonic trap the relation $E_0 \propto \langle r^2\rangle$ is
valid.  For condensates the trap determines the average properties. It
is then not very surprising that the numerical calculations of
$\langle r^2\rangle$ show that the interaction energy roughly is
proportional to the mean square radial difference between interacting
and non-interacting systems, i.e., $E-E_0 \propto \langle r^2\rangle
-\langle r^2\rangle_0$.  The similarity of these two sets of second
moments indicates that the corresponding wave functions also are
similar.  For weak interactions (very small scattering lengths) a
stationary many-body state can be approximated by a product of
single-particle amplitudes. However, stronger attraction between
particles must invoke other degrees of freedom like clusterization.
Then a simple single-particle description is not valid.

\subsection{Validity conditions for the models}

Validity criteria for our model and the mean-field approximation, both
for zero and finite-range interactions, can be compared for a
Bose-Einstein condensate where the wave function is located at
hyperradii $\rho \sim \sqrt{N}b\sub t$.  Accurate angular eigenvalues
in this region are crucial for a proper description.  If these
hyperradii are sufficiently large, i.e., $\rho \sim \sqrt{N}b\sub t >
N^{7/6}|a_s|$, the angular eigenvalue has reached its asymptotic value
where $\lambda \approx\lambda_\delta$.  This condition is equivalent
to $N|a_s|/b\sub t<N^{1/3}$ which is obeyed by stable condensates
where $N|a_s|/b\sub t < 0.5 <N^{1/3}$ \cite{boh98,rob01}.

The different models are valid if appropriately designed, i.e., our
model should reproduce the correct scattering length, whereas both the
zero and finite-range mean-field interactions should reproduce this
same correct scattering length but by using the Born approximation.
The interaction energies and sizes would all be similar for the states
corresponding to the condensate.  To make this comparison and reach
this conclusion we have to assume that the angular wave function is a
constant and that the hyperradial function is equivalent to the
single-particle product in mean-field computations, see
section~\ref{sec:wave-function}.  Otherwise the direct connection
between wave functions and their properties is impossible.  This
assumption about a specific form of the angular wave function is
similar to that of spherical Hartree-Fock computations for identical
fermions.

If we for a given average hyperradius $\bar\rho$, through
equation~(\ref{e3}), relate the mean-field average distance $\bar r$
by $\bar r\approx\bar\rho/\sqrt{N}$, then the density $n$ of the
system is given by
\begin{eqnarray}
  n
  \approx
  \frac{3}{4 \pi \bar r^3}
  \approx
  \frac{3N^{3/2}}{4\pi\bar\rho^3}
  \;.
  \label{eq:dens_rho}
\end{eqnarray}
The zero-range mean-field method is usually claimed to be valid for
condensates when $4 \pi n|a_s|^3/3\ll1$, see \cite{cow01,pit03}.  Then
the number of particles within a scattering volume $4 \pi |a_s|^3/3$
is on average much smaller than one.

On the other hand, in the zero-range asymptotic region of
$\bar\rho>N^{7/6}|a_s|$ we have $n \bar\rho^3>N^{7/2}n |a_s|^3$
immediately implying that $n|a_s|^3<1/N^2\ll1$, which means that the
system is very dilute and both zero and finite-range mean-field energy
is accurate.  For $\bar\rho<N^{7/6}|a_s|$ the large-distance
asymptotics are not valid and the zero-range mean-field description
breaks down. For $ N^{1/2}|a_s| \ll \bar\rho < N^{7/6}|a_s|$ or
equivalently $ 1/N^2 < n|a_s|^3 \ll 1$ the finite-range, but not the
zero-range, mean-field is valid.  For even smaller distances of
$\bar\rho < N^{1/2}|a_s|$ also finite-range mean-field becomes
invalid.

The present adiabatic hyperspherical method with two-body correlations
explicitly allowed in the form of the wave function is first of all
valid in the same region as the finite-range mean-field approximation,
i.e., for $N^{1/2}|a_s| < \bar\rho $, where correlations are expected to
be insignificant.  However, the validity range of the hyperspherical
method with two-body correlations incorporated extends to hyperradii
smaller than $N^{1/2}|a_s|$, where two-body correlations are
sufficient to describe the clusterizations.

When higher-order clusterizations occur, any method without
correlations higher than two-body breaks down.  The density when this
happens for this hyperspherical method is not easily derived.  The
lower limit is probably when the distance between two particles on
average equals the interaction range $b$, i.e., $N^{1/2} b < \bar\rho
$.  However, for nuclei with identical fermions the radius at
saturation is about $N^{1/3}b$ where the mean-field approximation is
very successful. This limit would then correspond to $N^{5/6}b <
\bar\rho $, but identical boson systems may allow even smaller
hyperradii.

In conclusion, the validity regions for the two-body correlated method
(hyperspherical), finite-range methods (finite-range mean-field and
hyperspherical), and the zero-range mean-field are estimated to be
\begin{eqnarray}
  \bar\rho > \sqrt N b 
  &\quad \textrm{for two-body correlated method,}
  \\
  \bar\rho > \sqrt N|a_s|
  &\quad \textrm{for finite-range methods,}
  \\
  \bar\rho > N^{7/6}|a_s| 
  &\quad \textrm{for finite- and zero-range methods.}
\end{eqnarray}
These relations can with equation~(\ref{eq:dens_rho}) be expressed via
the density
\begin{eqnarray}
  n|a_s|^3 < \Bigg(\frac{|a_s|}{b}\Bigg)^3
  &\quad \textrm{for two-body correlated method,}
  \\
  n|a_s|^3 < 1
  &\quad \textrm{for finite-range methods,}
  \\
  n|a_s|^3 < \frac{1}{N^2}
  &\quad \textrm{for finite- and zero-range methods.}
\end{eqnarray}
When the density is low, the three approximations are valid and the
energies are similar.  This assumes that the renormalization is
appropriate.  For higher densities the importance of correlations
increases and the mean-field approximations break down.  At even
higher density also two-body correlations are inadequate and the
particles want to exploit higher-order correlations.  In any case, the
wave functions can not be better than the Hilbert space they span, no
matter how precise the energy is computed.

\section{Summary and conclusion}

\label{sec:summary-conclusion}

The method of hyperspherical adiabatic expansion is briefly sketched
for a system of identical bosons.  The form of the wave function is
chosen as the $s$-waves in a partial wave expansion of the
Faddeev-Yakubovski{\u\i} cluster amplitudes.  This restriction is
expected to be accurate for large distances and dilute systems.  We
relate to the Jastrow ansatz designed to deal with correlations in
rather dense systems.  We discuss the theoretical connections between
these approaches and the mean-field approximation both with zero and
finite-range interactions.

The angular eigenvalues in the hyperspherical adiabatic expansion
appear as crucial ingredients in the radial potentials.  We use the
analytic expressions recently parametrized to reproduce the results of
full numerical computations.  We first discuss the general properties
of these eigenvalues as functions of hyperradius for arbitrary
particle number and arbitrary scattering length.  The large-distance
behaviour corresponding to the zero-range mean-field result is
obtained.

The radial potential has a minimum at large distance when particle
number times scattering length divided by trap length is less than
about 0.5.  The wave function of the condensate is located in this
minimum. In addition, for sufficiently large scattering lengths an
intermediate region appears with a radial potential decreasing
inversely proportional to the square of the hyperradius.  This region
supports the many-body Efimov-like states.  At much smaller distances
a pronounced attractive pocket is present when the two-body potential
is attractive.  We give analytical estimates of the number of bound
states located in these different regions.  We then discuss the decay
properties eventually arising from recombination processes.  In
particular the highest-lying Efimov-like states located at large
distances recombine corresponding to widths much smaller than the
level spacing.  These peculiar states could then leave observable
traces.

Finally, we discussed the connection between this work and the
mean-field approximation.  We first emphasized that the effective
two-body interactions must be related to the Hilbert space for the
wave function.  We specify the necessary renormalization for the
mean-field restriction.  Numerical comparison for energies and radii
are then presented.  The validity conditions for the models are
discussed and expressed as regions in hyperradius.  These regions
increase from zero via finite-range mean-field approximation to the
hyperspherical adiabatic expansion method.  Most of the results are
independent of the structure of the two-body interaction.  The
conclusions are derived in terms of scattering length, number of
particles, external field frequency and occasionally the effective
range of the two-body potential.
\\~\\


\end{document}